\documentclass{PoS}
\usepackage{natbib}
\bibpunct{(}{)}{;}{a}{}{,}

\title{Radio-loud AGN-jet morphology and polarization: the role of ultra-high resolution radio surveys}

\ShortTitle{Radio-loud AGN-jet morphology and polarization: the role of ultra-high resolution radio surveys}

\author{Iv\'an Agudo\\ 
        Instituto de Astrof\'{\i}sica de Andaluc\'{\i}a (CSIC),  
                 Apartado 3004, E-18080 Granada, Spain\\
        E-mail: \email{iagudo@iaa.es}}

\abstract{A review of the current challenges for the understanding of the physics of extragalactic radio jets from supermassive black holes is presented.
Also, a prospect is given about how both very long baseline interferometry (VLBI) and polarimetric observations help us to understand their physics and the ones of their magnetic fields. 
This paper focuses on the impact that previous ultra-high resolution polarimetric surveys have had on our current knowledge on radio loud AGN. 
It first concentrates on the expectations about the improvement that the Square Kilometer Array, used as a super-sensitive VLBI station on existing and future very long baseline interferometric arrays, will provide in terms of access to new source classes, and in terms of a deeper portion of the Universe than currently accessible. 
A series of key new radio loud AGN science fields, to be opened by the Square Kilometer Array, are outlined together with a collection of preliminary ideas about possible programs of interest to deep in such fields.
The complementary aspects of future super-sensitive VLBI observations as compared to deep continuum surveys with the Square Kilometer Array and its precursors are also outlined here.}

\FullConference{EXTRA-RADSUR2015 (*)\\
		20--23 October 2015\\
		Bologna, Italy

                \bigskip
                \hrule
                \bigskip

                \textnormal{(*) This conference has been organized
                  with the support of the Ministry of Foreign Affairs
                  and International Cooperation, Directorate General
                  for the Country Promotion (Bilateral Grant Agreement
                  ZA14GR02 - Mapping the Universe on the Pathway to
                  SKA)}}

\begin{document}

\section{Introduction}
Relativistic jets in radio-loud active galactic nuclei (AGN) are among the most energetic objects known so far, and among the brighter radio emitting objects in the sky.
The jets from this kind of objects can emit prominent and extremely variable polarized synchrotron radiation at all spectral bands from radio to $\gamma$-rays.
An inverse Compton component of emission is also possible, perhaps even prominent, at high energies (from X-rays to $\gamma$-rays) in this kind of objects.
Bright radio-loud AGN are ubiquitous in the sky, and therefore are expected to appear in nearly any new extragalactic radio survey (either deep or not).
Since they are highly luminous, they can be detected up to large cosmological distances, which opens the opportunity for the study of the astrophysics of yet unexplored ages of the Universe.
For this,  future ultra-high sensitivity radio-surveys \citep[see e.g.,][]{Agudo:2015p24803} such as those starting to be planned for the main scientific cases driving the Square Kilometer Array (SKA) project \citep{Braun:2015p24804} will be instrumental.
Moreover, the polarized nature of the radio emission in radio-loud AGN jets makes them ideal background sources for studies of the large scale magnetic field of the Universe \citep{Gaensler:2015p24805, JohnstonHollitt:2015p24807, Taylor:2015p24806}. 

The jet formation, collimation and acceleration processes are among the most relevant but most poorly known problems on current jet astrophysics. 
We have recently started to understand these processes with unprecedented detail from the numerical point of view.
However, we are still lacking most information and confirmations from the observational side.
We know that the essential ingredients to produce relativistic jets in AGN are the large gravitational potential of the central rotating supermassive black hole (SMBH), the surrounding material from the rotating accretion disk, and its co-rotating magnetic field.
This makes jet formation studies a powerful tool to probe the environment of SMBH, and the physics of high-energy plasmas, and their magnetic fields.
Therefore, ultimately, AGN relativistic jet studies aim to attack fundamental problems in physics.
Obviously, because of the key role of magnetic fields in the formation and evolution of relativistic jets, polarization studies of these objects have the potential to provide key information to understand the processes involved.

During the latest years theoretical argumentation has made clear that after formation, for jets to be efficiently collimated, a mechanisms would need to operate producing stratified jets in both composition (outer electron-proton wind, inner electron-positron) and speeds.
In this case, jets can produce an ultrafast and light relativistic jet spine mainly composed of electron-positron plasma, plus a slower and heavy sheath mainly composed of electrons and protons. 
This issue about jet composition is another of the current great challenges that needs to be solved from the observational perspective.
There are many chances that future instruments like SKA, with multiple high-frequency, and high-precision, full polarization capabilities, will have the opportunity to make a big step forward in this field.
This will benefit from the use of the ultra-high angular resolution feature of very long baseline interferometry (VLBI) in combination with SKA as an additional super-sensitive station \citep[SKA-VLBI,][]{Paragi:2015p25183}.

After their formation in the innermost scales of the AGN, powerful relativistic jets can keep well collimated up to scales that are far larger than the size of their host galaxies.
But there is still a relevant problem regarding that observational fact: how can jets radiate such enormous amount of power (up to $\sim10^{48}$\,erg/s) from the sub--pc scales, up to scales of hundreds of kpc (or even larger)?
For the case of AGN jets, this is related to an additional fundamental question: how and at what distance from the central engine Poynting-flux dominated outflows are converted into kinetic dominated jets?
These two fundamental questions are related to a general lack of our current understanding of the particle acceleration processes.
Among these processes, the three more feasible ones are diffusive acceleration at the fronts of strong hydrodynamic (HD) shocks, magneto-hydrodynamic (MHD) reconnection, and stochastic MHD interactions of particles with magnetic turbulence.
In all these three cases, the resulting non-thermal particle spectra (that can be measured) depend strongly on the underlying physical conditions, which provides us with a tool to constraint the kind of acceleration mechanism at work from AGN jet observations.
All these processes take place in the innermost regions of AGN relativistic jets, at the pc and sub-pc scales, as does the jet formation, collimation and acceleration process.
This region, the most compact and therefore the most poorly understood one, is in general only visible through the ultra-high angular resolution VLBI, which is therefore key to understand such relevant questions outlined above.

\section{Capabilities of current VLBI facilities}

\begin{figure}
    \centering
    \includegraphics[width=1.0\textwidth]{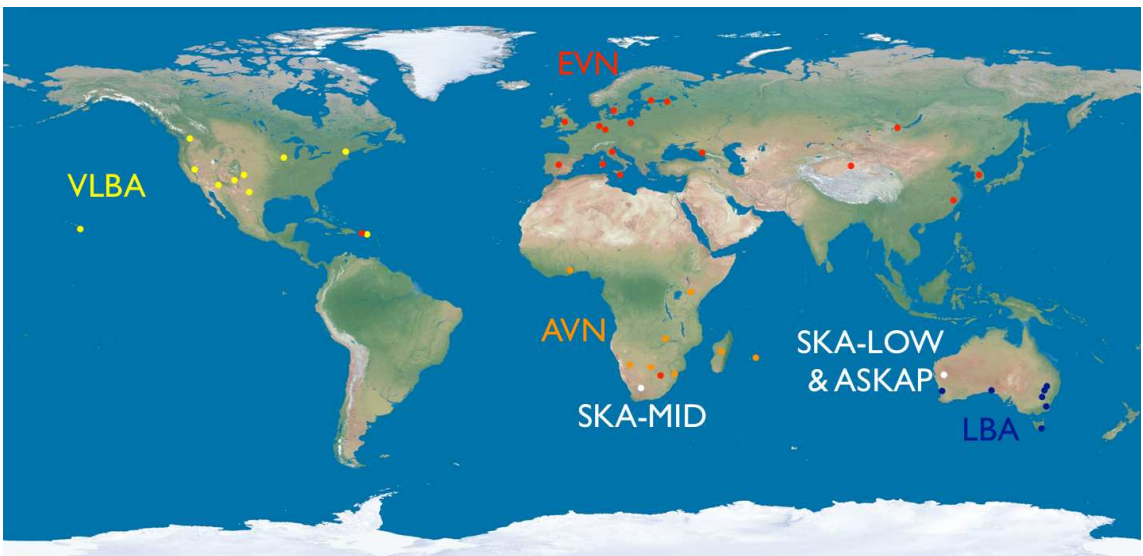}
    \caption{Geographical distribution of current and future centimeter wavelength VLBI arrays.}
    \label{vlbimap}
\end{figure}

A selection of the most relevant VLBI arrays for the kind of studies outlined in this paper contains the European VLBI Network\footnote{\tt http://www.evlbi.org} (EVN, mainly located in the Euro-Asiatic continent but also including stations in South Africa and Puerto Rico), the Very Long Baseline Array\footnote{\tt https://public.nrao.edu/telescopes/vlba} (VLBA, in the USA), and the Long Baseline Array\footnote{\tt http://www.atnf.csiro.au/vlbi} (LBA, in Australia), see Fig~\ref{vlbimap}.
Besides combination between these three arrays (which are certainly possible), these are the three largest ground interferometers, and therefore provide the finest angular resolutions among all astronomical facilities on Earth (up to $\sim0.15$\,milli-arcseconds at 43\,GHz [7\,mm]).
However, the current record on astronomical angular resolution ($\sim14\,\mu$as) comes through VLBI arrays including an orbiting station, i.e. the RadioAstron\footnote{\tt http://www.asc.rssi.ru/radioastron} orbiting antenna \citep{Kardashev:2012p24991}, that still operates successfully and releases pioneering results (see Kovalev et al., these proceedings).

Typical sensitivities of $10-15\,\mu$Jy/beam (for 1h integration time) are routinely achieved at the EVN, the most-sensitive VLBI array so far.
However, combining the current standard VLBI arrays with a number of very sensitive stations operating in the northern hemisphere (e.g. EVN involving the Arecibo 300\,m Telescope, the 70\,m antenna in Robledo de Chavela, and a number of new 65\,m stations of the EVN [e.g. Sardinia, Italy and Tian Ma, China]; or the VLBA plus the 100m Green Bank Telescope, the phased JVLA, and the Effelsberg 100m Telescope), the VLBI sensitivity can be improved by a factor $>3$, i.e. up to $\sim3\,\mu$Jy/beam/h.
The organization of these high-sensitivity arrays is not a trivial task though, and the standard sensitivities of VLBI observations do not usually reach such great levels.
Furthermore, most of the southern sky is not currently accessible by high-sensitivity VLBI because of the geographical distribution on the northern hemisphere of such high-sensitivity arrays.

\section{Previous VLBI surveys and some of their achievements}

\subsection{Continuum VLBI surveys}

Most of the large and early total-flux VLBI surveys of radio-loud AGN focused on the study of the structure and compactness of flat-spectrum radio jets. 
The observations were performed at 5 GHz with the VLBA, and even with the VLBI Space Observatory Program\footnote{\tt http://www.vsop.isas.jaxa.jp/top.html} \citep[VSOP,][]{Hirabayashi:2000p24920}, a project operated by Japan in the late nineties and beginning of the 2000's that pioneered VLBI with an orbiting antenna. 
In the following lines, a (probably biased and certainly incomplete) set of reference programs is listed.
These programs made relevant technical and/or scientific contributions to the study of radio-loud AGN jets since mid nineties, when the main modern VLBI arrays started operating.

The Caltech--Jodrell Bank Survey \citep[CBS, see ][and references therein]{Taylor:1994p25062,Polatidis:1995p25076}, produced VLBA total-flux images at both 5 and 1.7\,GHz for a complete flux density limited sample of ~300 bright ($S>350$\,mJy at $\sim5$\,GHz) and flat spectrum AGN. 
This was the first milli-arcsecond scale morphological classification of a very large sample of radio loud AGN sources.

A few years later, in June 1996, the VLBA Pre-Launch VSOP Survey \citep{Fomalont:2000p4880}, in preparation for a big AGN jet survey with VSOP, made imaging observations of 374 strong flat-spectrum radio sources north of declination $-44^{\circ}$ using the VLBA at 5\,GHz. 
Among all observed sources, a large fraction of them were selected for the VSOP survey \citep{Hirabayashi:2000p25096,Lovell:2004p25104,Scott:2004p17582,Horiuchi:2004p17583}.
The VSOP survey observed 294 flat-spectrum AGN brighter than $1$\,Jy at 5\,GHz.
A significant fraction of the observed sample (54\,\%) was found to display brightness temperatures of the innermost compact cores $T_{b}>10^{12}$\,K (with an observed maximum of $T_{b}=1.2\times10^{13}$\,K).
This implied the confirmation of the old standing problem of the very-high brightness temperature of flat-spectrum radio-loud AGN that requires a too large amount of relativistic Doppler beaming, not compatible with the observed proper motions in a small number of sources.
The problem is still not quite solved and it is matter of intense debate.
A detailed update on the status of the problem is provided by Kovalev et al. (these proceedings), where a comprehensive space VLBI survey of AGN involving the RadioAstron orbiting station to attack this problem is also presented. 

When VLBI arrays started to have enough sensitivity, and the computational power of correlators started to be suited for that, the first attempts to image the deep radio sky on wide areas started to come. 
\citet[][and references therein, see also Radcliffe et al. these proceedings]{Garrett:2001p25158,Chi:2013p25138} have been pioneers on very-deep and wide field VLBI imaging of compact structures at large redshifts.
This observing method is extremely useful to identify the radio AGN population on a sample of weak galaxies; for the Hubble Deep Field-North in the case of the above mentioned references.
In particular, the observations by \citet{Chi:2013p25138}, made with a Global array at 1.4\,GHz, were deeper and wider than any other previous VLBI observations of such field (achieving a minimum rms$=7.3\,\mu$Jy/beam at 4\,milli-arcsecond resolution on a area of the sky extended by $\sim200$\,arcminutes$^2$).
Although still limited by small number statistics, \citet{Chi:2013p25138} found that 25\,\% of star forming galaxies (SFG) contain faint AGN.
\citet{Garrett:2001p25158} and \citet{Chi:2013p25138} did not only show the potential of deep and wide field VLBI observations for AGN-SFG studies, but also have demonstrated a tremendously useful application of VLBI to support population studies from deep surveys with SKA, which addition will greatly enhance the sensitivity.

The mJIVE-20 project \citep{Deller:2014p25139} is, in a sense, a massive extension of the pioneering work by \citet{Garrett:2001p25158} and \citet{Chi:2013p25138}.
This program has imaged hundreds of different wide fields and tenths of thousands of sources contained on them down to the milli-Jansky level.
This is the largest 20\,cm VLBI imaging survey so far. 
The project, developed at the VLBA, took advantage of the multi-phase center correlation capability recently developed in most VLBI software correlators to image 25973 radio sources from where thousands of new VLBI sources down to $\sim1$\,mJy were detected.

\subsection{Polarimetric VLBI surveys}
\label{polsurv}

In a pioneering work, \citet{Zavala:2004p138} presented the first survey for mapping the Faraday rotation measure along a big number of (40) different jet sources through VLBA observations made at seven different frequencies between 8.1 and 15.2\,GHz. 
Their source sample included quasars, radio galaxies, and BL~Lacertae type objects (BL~Lacs), that were observed on a single epoch each.
Their results show core rotation measures ($RM$) of both quasars and BL Lacs ranging from a few hundreds to a few thousands of rad/m$^2$, with downstream jet features showing smaller $RM\lesssim500$\,rad/m$^2$, and radio galaxies showing in general much larger jet rotation measures (with $RM$ in the range of a few hundreds to $\sim10^5$\,rad/m$^2$) than quasars and BL Lacs.
Radio galaxy cores are also generally depolarized, which is consistent with the idea of radio galaxy jets lying close to the plane of the sky, whereas jets in quasars and BL Lacs  point close to the line of sight.
The properties of the Faraday screen were not clear from the early work of \citet[][and references therein]{Zavala:2004p138}, and they are still under debate.

On more recent work, the group by \cite{Gabuzda:2015p25146}, through a set of VLBA imaging observations at a maximum of different 7 frequencies (typically from $\sim5$ to $\sim15$\,GHz) have shown the existence of systematic Faraday rotation gradients across a significant number of parsec scale jets.
This is claimed to be in rather good agreement with the existence of helical magnetic fields in the innermost regions of jets. 
This is not a definitive prove though, but it is a potential relevant piece of evidence for the confirmation of jet formation theories. 
The recent results presented by the MOJAVE\footnote{\tt www.astro.purdue.edu/MOJAVE} (Monitoring Of Jets in Active galactic nuclei with VLBA Experiments) team \citep{Lister:2009p5316}, i.e. \citet{Hovatta:2012p18733}, and earlier work made on individual sources \cite[e.g.][]{Zavala:2005p399,2008ApJ...682..798A,Gomez:2008p30,2008ApJ...675...79A}, also agree with the claims by \cite{Gabuzda:2015p25146} for helical magnetic fields threading the jets from radio-loud AGN.

\begin{figure}
    \centering
    \includegraphics[width=0.9\textwidth]{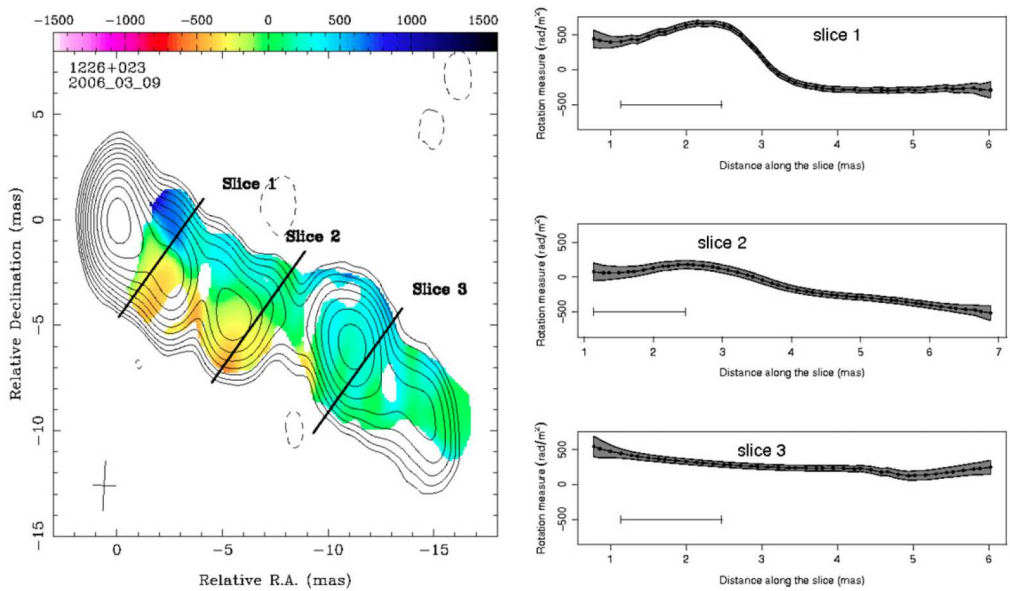}
    \caption{Rotation measure map of 3C 273 at milli-arcsecond resolution as obtained with the VLBA by the MOJAVE team. Reproduced from \citet{Hovatta:2012p18733}.}
    \label{3C273}
\end{figure}

The contribution of the MOJAVE monitoring survey to the knowledge of the radio loud AGN population on the parsec scales is indeed unprecedented, and certainly not limited to Faraday rotation studies.
MOJAVE uses the VLBA to observe the 15\,GHz total flux and polarization (linear and circular) structure of different source samples containing more than 200 sources.
The program monitors sources (since 2002) with a time sampling up to one month, for sources with time scales of variability of that order.
This MOJAVE survey has made important contributions on defining the kinematic and geometric jet parameters of hundreds of jet sources  \citep{Lister:2009p5316,Lister:2009p6166,Pushkarev:2009p9412,Homan:2009p9257,Savolainen:2010p11632}, on defining the jet regions where collimation and acceleration take place \citep{Homan:2015p25178}, and on characterizing the full polarization properties of relativistic AGN radio--jets \citep{Homan:2006p238}.
The latter includes circular polarization studies, that are of extreme relevance to make accurate fits and estimates of the actual macro-- and micro--physics determining the observational properties of these sources \citep{Wardle:2003p8612,Homan:2009p6162}.

An also extremely useful AGN-jet VLBI monitoring survey is the one maintained by the Boston University (BU) Blazar Group since 2007; the year when the launch of the \emph{Fermi} $\gamma$-ray Space Telescope (still in operation) was expected.
The BU Blazar Monitoring Program\footnote{\tt https://www.bu.edu/blazars/VLBAproject.html} monitors a set of $\sim40$ of the brighter blazars and radio galaxies with a time sampling of about one month with the VLBA both in total flux and with polarimetric sensitivity. 
One of the nicest properties of this monitoring survey is that it observes at very high frequencies ($43$\,GHz), and therefore it images essentially the innermost visible regions of blazar jets, were the sources are still not affected by synchrotron self-absorption at these high frequencies.
This allows for both the total flux and the polarization properties of the sources to be imaged without of the influence of opacity effects, that would greatly distort our view on these inner regions.
Therefore, the BU Blazar Monitoring Program is particularly well suited, and it has been extremely successful, on its combination with comprehensive sets of multi--spectral--range data (with polarimetry whenever possible), to identify and characterize the different emitting regions along the spectrum (up to $\gamma$-rays) and along the jet.
The polarimetric component of the VLBI observations is critical here in many cases where the identification of polarized moving features in the jet with events happening in the optical is performed through the matching polarization angles of contemporaneous flares.
This program has been demonstrated to be particularly useful in constraining, or even determining, the possible high-energy emission mechanisms of blazar jets \citep[e.g.][]{Marscher:2008p15675,Marscher:2010p11374,Jorstad:2010p11830,Jorstad:2013p21321,Agudo:2011p14707,Agudo:2011p15946} which are among the current challenges for high--energy astrophysics.

\section{A perspective for future VLBI surveys of radio-loud AGN jets}

\subsection{New ultra-sensitive VLBI arrays}

To have a realistic view of how the ultra-high resolution surveys of radio-loud AGN jets that are expected to come will impact the astrophysics of these objects, it is important to have an idea of the instrumental improvements that are planned for the near future.
Besides the continuous development of digital VLBI backends at virtually all current interferometers, and therefore the resulting increase of observing bandwidths and sensitivity, the EVN is, among the current VLBI arrays, the one that profits from a faster addition of new high-sensitivity stations of the class of 65\,m antennas (i.e. the 65\,m Sardinia Radio Telescope [Italy], and the Tian Ma 65\,m Radio Telescope [China], see above).
Besides that, the greatest jump in sensitivity that is expected in VLBI observations comes with the advent of the SKA.
In its Phase 1 (2020--2030), the South African component of SKA \citep[i.e. SKA-MID,][]{Braun:2015p24804} is expected to participate on global arrays for VLBI interferometers where the central core (with $\sim80$\,\% of the sensitivity of the entire SKA-MID array) will act as a super-sensitive station.
This station will govern the sensitivity of the entire global array \citep[see][]{Paragi:2015p25183}.
For the Phase 2 of SKA (from 2030), a new dramatic increase on sensitivity, image fidelity, and dynamic range are expected through the use of VLBI observations with a homogeneous array of SKA-MID stations even involving intercontinental baselines.

For SKA1-MID the concept of the African VLBI Network\footnote{http://ska.ac.za/avn} (AVN) is starting to be developed.
This project plans to modify existing, but redundant, dishes previously used for satellite telecommunication into radio telescopes and VLBI stations in a number of african countries, i.e. Botswana, Ghana, Kenya, Madagascar, Mauritius, Mozambique, Namibia, and Zambia.
Since there are no other VLBI stations to cover baselines from a few hundreds to a few thousands of km, the AVN would give excellent support to a sensitive VLBI array involving SKA1-MID, not to mention  the great value of having a dedicated VLBI array ($\sim100$\,\% of the time) in the southern hemisphere.
Either with or without the AVN, the EVN (covering a similar range of longitudes as the AVN) will provide support to VLBI observations with SKA-MID.
Of course, having the additional participation of the Australia SKA Pathfinder (ASKAP), the LBA, the VLBA, and any other potential space station (or stations) at that time, would be much more than desirable, but most of these options would provide only a partial coverage to VLBI observations with SKA-MID.
Even without these options, using the phased SKA1-MID together with the AVN and the EVN it will be possible to reach ultra-high angular resolutions up to a few tenths of a milliarcsecond at $\mu$Jy rms levels (as well as astrometric precisions up to a few micro/arcseconds only).

\subsection{New science with improved VLBI capabilities}
\label{sci}

One of the most obvious and early applications of VLBI observations involving SKA will be the identification of source populations from previous SKA alone surveys, in particular to distinguish radio AGN from star forming galaxies \citep[e.g.,][]{Agudo:2015p24803}.
It is obvious that linear polarization and spectral index information from such SKA surveys will be a useful tool for that, as well as, optical data providing either photometric or spectroscopic redshifts.
Moreover, VLBI observations will provide the definitive confirmation of the AGN nature of those sources detected with high VLBI brightness temperature exceeding those expected for start forming regions.
Polarimetric VLBI observations will also be critical for this kind of work, and they come for free in terms of additional observing time.

The coming high-sensitivity capabilities of VLBI observations open a completely new opportunity to study the evolution of AGN jets along cosmic time.
This is almost an entire new field to explore that needs to answer relevant questions as: when where the first AGN formed?, how did they form?, how were their environments and those of their SMBH and accretion systems? or how did they evolve with time up to $z\sim0$?
To attack these questions, VLBI observations will be instrumental not only on the identification of sources, but also on the characterization of the properties of AGN along cosmic time.

In particular, regarding the detection and possible characterization of the first AGN SMBH, their appearance is highly speculative, and there are lots of uncertainties regarding predictions so far \citep{Rees:1978p25210,Heger:2001p25218,Shapiro:2005p25226}.
However, simulations predict that dense clouds might be able to form massive compact objects containing accretion disks \citep{Schleicher:2013p25193} that would naturally favor the formation of small-scale jet embryos frustrated by the dense environment (a far-away Universe version of nearby gigahertz peaked radio sources, GPS).
This is actually favorable in terms of the detection of these kind of objects \citep{Afonso:2015p25184}, because in these cases the radio emission is not dramatically affected by IC looses by the dense cosmic microwave background (CMB). 
This means that there are possibilities to detect these sources at 100-600 MHz on the 0.1\,mJy at arcsecond resolution, as proposed by \citet{Falcke:2004p25189}.
Again the high resolution of future ultra-deep VLBI observations involving SKA-LOW \citep[the low frequency component of SKA in Australia,][]{Braun:2015p24804} and/or SKA-MID (observing at low frequencies) will be essential to identify and characterize this kind of sources.

An additional application of very--high sensitivity and very--high resolution radio observations involving future instruments is the detection of radio-loud AGN at high redshifts at $z\gtrsim8$.
\citet{Ghisellini:2014p22313} have made a great advance on this topic through simulations of powerful relativistically-beamed radio-loud AGN (i.e. blazars) at high redshifts.
They show that the radio synchrotron component of the sources fade strongly with redshift by more than 6 orders of magnitude in luminosity.
This is a consequence of the density of the CMB at high redshifts, that produces strong high-energy radiative losses. 
This is responsible for the strong reduction of the relativistic electron population that radiates the synchrotron radio emission, and therefore the radio spectrum fades off \citep[see also][]{Afonso:2015p25184}.
However, \citet{Ghisellini:2014p22313} have provided excellent news on which regards to the potential detection of radio-loud AGN at $z\sim8$ by making use of the unprecedented sensitivity of the SKA-MID even on its early Phase 1 \citep[see also][]{Agudo:2015p24803}.
According to \citet{Ghisellini:2014p22313}, SKA1 will be able to detect the radio faint emission from most of the bright blazar population up to $z\approx8$, therefore being able to observe many more jetted sources than predicted on the basis of current deep radio surveys.
Indeed, this will not only allow cosmological studies of radio-loud AGN since the epoch of the first AGN, but also cosmological studies of the CMB itself.
This will be possible by studying the impact of the CMB on the spectral energy distribution (SED) of blazar jets, as proposed by \citet{Ghisellini:2014p22313}. 
These kind of studies would need to use data from wide-area and deep surveys at the broadest possible frequency range (both in all SKA bands and at higher frequencies/energies) to characterize the SED of the sources.
VLBI observations will also be essential to make a characterization of both the first AGN jets, and their dense environments and magnetic fields.

Tidal disruption events (TDE) of stars in the surroundings of SMBH, as a particular case of AGN, can also be studied adequately by the same methods as those usually used to study the most transient class of AGN (i.e. blazars, see Section~\ref{polsurv}).
In particular, it is expected that at least hundreds of new TDE will be discovered by SKA, which will give the opportunity to trigger multi-spectral-range campaigns, hopefully also involving multi-frequency polarimetric VLBI.
If so, the ultra-high resolution images provided by VLBI will be essential to characterize the expansion rate of the newly formed jet \citep{Donnarumma:2015p24802} in the same way as it is routinely done on extreme flaring events on blazars \citep[e.g.][]{Marscher:2008p15675,Marscher:2010p11374,Jorstad:2010p11830,Jorstad:2013p21321,Agudo:2011p14707,Agudo:2011p15946}. 

\section{A collection of desirable SKA-VLBI surveys for radio-loud AGN-jet research}

A few preliminary (personal, and therefore perhaps biased) ideas about possible VLBI surveys and programs to develop with the SKA1 (and its precursors) regarding radio-loud AGN-jet research are summarized here.
But first, it is worth mentioning that none of the concept VLBI programs outlined below will be possible without the information provided by previous wide and/or deep SKA surveys that are being planned for the key science of SKA1. 
The main source properties (i.e. position, extended morphology, radio spectrum, and linear and circular polarization properties) that will lead to their selection for further follow up studies with ultra-high resolution will come from previous arcsecond resolution SKA surveys, which are therefore instrumental for any further VLBI follow up.

As mentioned above, a wide-field, targeted, and deep SKA-VLBI survey to discriminate and identify source classes (i.e. radio AGN vs. SFG) would be the first and most direct application of VLBI observations.
For sources at $z>1$, SKA-MID alone will not reach angular resolution enough for this kind of studies, and SKA-VLBI will be essential.
SKA-VLBI observations, designed from previous SKA surveys, will be an instrumental tool to understand and interpret the results of the first SKA continuum surveys.
These VLBI observations will be better performed at L-band (1-2\,GHz), which is a classical VLBI band where most radio observatories participating on global VLBI arrays have available receivers.
Polarization sensitivity on these VLBI observations will be more than useful for source discrimination at no cost in terms of observing time, which is true for any other VLBI program. 

An ultra-deep SKA-VLBI targeted follow up of potential high-redshift ($z\gtrsim5$) sources would be of top interest to study both the blazar-like early population of radio-loud AGN, and the environments of early SMBH through the signature imprinted by their associated GPS-like jets (see Section~\ref{sci}).
This observing program could, in principle, be combined on a commensal way, with the one outlined above for radio AGN vs. SFG discrimination if the observations for high-redshift sources are planned to be performed as deep as possible.
Because of the large redshift nature of the source populations, observations at the low end of the VLBI frequency window, i.e. the P-band (300-800\,MHz), would be desirable as well.
This implies an additional complication for observations, though, given the restricted number of available stations operating at such low frequencies.

Finally, a target-of-opportunity-like SKA-VLBI monitoring program of transient phenomena such as TDE and extraordinary blazar flares would be of great interest to explore the still unknown innermost regions of relativistic jets.
These inner regions are those where the jet formation, collimation, and acceleration processes still have a relevant influence on the observable properties of objects, and therefore their study should shed new light on those processes.
Full-polarization multi-frequency observations with a minimum of 3 well separated frequency bands in the 5\,GHz to 24\,GHz range (to allow for high rotation measure studies), and with adequate time sampling for every kind of event, would be an ideal choice for this kind of program in order to allow for the highest angular resolutions.

\section{Summary and concluding remarks}

The relevance of the current challenges on the physics of radio-loud AGN jets (some of them long-term problems in the field) has been shown.
Although the advent of the next generation of radio astronomical facilities (i.e. the SKA) is already in its path, the addition of an ultra-high resolution component and the guarantee of very--high--precision full--polarization capabilities will enhance the science outcome of the SKA.
It is anyhow clear that the base of any follow up ultra-high resolution study should be the information obtained from previous  arcsecond resolution SKA surveys, that will provide the first new insights on the current astrophysical problems on radio-loud AGN.
It has been shown how the past and recent VLBI surveys have left a relevant imprint on our knowledge of the jet phenomenon from SMBH in AGN, and how polarimetric programs have greatly enhanced such advances.
The future is even more promising, once ultra-high sensitivity observations can be performed with the SKA. 
Clearly, high precision polarimetry (including circular polarization), and VLBI observations have a relevant role on the new era of radio astronomy.
Both of them will open new windows for the study of new scenarios.
These include both proto-AGN jet systems at $z\gtrsim8$ and the first blazars in the Universe, as well as the study of their evolution until $z\sim0$ (to be searched through the deepest low frequency VLBI observations).
Long standing problems like the jet formation, collimation, and acceleration processes; and the determination of the jet composition will also have a relevant advance from multi-(high)frequency full-polarimetric observations.
It has been shown that at the times of SKA, a sensitive VLBI component with good polarimetric capabilities in the southern hemisphere will be an essential tool.
Because of this, it would be desirable for MeerKat (the precursor of SKA-MID in South Africa) to establish a plan to add a VLBI component to its collection of observing modes (perhaps based on the AVN concept plus additions from the EVN) in order to have the opportunity of enjoying the nice capabilities of ultra-high resolution right from the beginning of its operations.

\section*{Acknowledgements}
     The author acknowledges the organizers of this conference for their great hospitality and their excellent job on the scientific and logistical organization of the meeting.
     Zsolt Paragi and Roger Deane are acknowledged for providing useful comments that helped to improve this manuscript. 
     The author acknowledges support by a Ram\'on y Cajal grant of the Ministerio de Econom\'ia y Competitividad (MINECO) of Spain.
 

\end{document}